\documentclass[11pt,a4paper,epsf,epsfig,psfrag]{article}
\usepackage{jheppub}
\usepackage{amsmath,graphicx,amsfonts, mathrsfs,amssymb}
\usepackage{amsmath,epsfig}
\usepackage{amssymb,amsfonts}
\usepackage{latexsym}
\usepackage{epsfig}
\usepackage{pdfsync}

\newbox\pippobox

\title{Two-gluon and trigluon glueballs from dynamical holography QCD}
\author{Yidian Chen$^{a}$,  Mei Huang$^{a,b}$\\
$^{a}$ {Institute of High Energy Physics, Chinese Academy of Sciences, Beijing, China} \\
$^{b}$ {Theoretical Physics Center for Science Facilities, Chinese Academy of Sciences, Beijing, China} \\}

\abstract{We study the scalar, vector and tensor two-gluon and trigluon glueball spectra in the framework of 5-dimension
dynamical holographic QCD model, where the metric structure is deformed self-consistently by the dilaton field. For comparison,
the glueball spectra are also calculated in the hard-wall and soft-wall holographic QCD models. In order to distinguish glueballs
with even and odd parities, we introduce the positive and negative coupling between the dilaton field and glueballs, and for higher
spin glueballs, we introduce a deformed 5-dimension mass. With this set-up, there is only one free parameter
from the quadratic dilaton profile in the dynamical holographic QCD model, which is fixed by the scalar glueball spectra.  
It is found that the two-gluon glueball spectra produced in the dynamical holographic QCD model are in good agreement 
with lattice data. Among six trigluon glueballs, the produced masses for $1^{\pm -}$ and $2^{--}$ are in good agreement 
with lattice data, and the produced masses for $0^{--}$, $0^{+-}$ and $2^{+-}$ are around 1.5 {\rm GeV} lighter than 
lattice results. This result might indicate that the three trigluon glueballs of $0^{--}$, $0^{+-}$ and $2^{+-}$ are dominated 
by three-gluon condensate contribution.}


\begin{document}
\maketitle

\newpage

\section{Introduction}

Quantum chromodynamics (QCD) is accepted as the fundamental theory of describing
strong interaction. In the high energy regime,  QCD has the property of asymptotic freedom
and perturbative QCD calculations have been tested with high precision. However, in the low
energy regime, the nonperturbative aspect related to QCD vacuum properties and hadron
spectra remains as outstanding challenge. The nonabelian feature of QCD makes
it possible to form bound states of gauge bosons, i.e. glueballs (gg, ggg, etc. ) \cite{Glueball-first}.
The gauge field plays a more important dynamical role in glueballs than that in the standard hadrons,
therefore study particles like glueballs offers a good opportunity of understanding nonperturbative
aspects of QCD.

The glueball spectrum has attracted much attention more than three decades \cite{Glueball-first},
and it has been widely investigated by using various non-perturbative methods. For example, from first principles
calculation by using lattice QCD \cite{Morningstar:1999rf,Lucini:2001ej,Meyer:2004gx,Chen:2005mg,Gregory:2012hu},
by using flux tube model \cite{Isgur:1984bm} as well as by using QCD sum rules \cite{Huang:1998wj,Qiao:2014,Qiao:2015iea}.
For more information, please refer to review papers \cite{Glueball-Review}.

The discovery of the gravity/gauge duality, or anti-de Sitter/conformal field theory (AdS/CFT) correspondence
\cite{Maldacena:1997re,Gubser:1998bc,Witten:1998qj}
offers a new possibility to tackle the difficulty of strongly coupled gauge theories, for reviews see Ref. \cite{Reviews}.
In recent decades, many efforts have been invested from both top-down and bottum-up approaches in examining
nonperturbative QCD properties, e.g.,  QCD equation of state, phase transitions,
fluid properties of quark-gluon plasma \cite{Kovtun:2004de}, meson spectra  \cite{EKSS2005,Karch:2006pv,mesons}, baryon spectra \cite{baryons},
as well as in the glueball sector \cite{glueball-topdown,glueball-AdS-operator,glueball-bottomup,Colangelo:2007pt,Forkel:2007ru}.
It is expected that the holography approach can shed some light on our understanding of the nonperturbative aspects of QCD.

QCD is a non-conformal gauge theory, and the Sakai-Sugimoto (SS) model \cite{SS} is one of the most successful
non-conformal top-down holographic QCD models. The glueball spectra in the Sakai-Sugimoto model has been
investigated in literatures, see Ref. \cite{glueball-SS}. Glueballs have also been widely studied by using the
bottom-up approach \cite{glueball-bottomup}, where most studies are based on hard-wall \cite{EKSS2005}
and soft-wall holographic QCD models \cite{Karch:2006pv} with the conformal $AdS_5$ background metric.

A successful holographic QCD model should grasp two main features of nonperturbative QCD properties,
i.e. the spontaneous chiral symmetry breaking and color charge confinement. The dynamical holographic QCD model
(DhQCD) which can describe both chiral symmetry breaking and confinement
has been constructed in Ref. \cite{Li:2012ay,Li:2013oda,Li:2013xpa}. In this model, the gluon dynamics background is
determined by the coupling between the graviton and the dilaton field $\Phi(z)$, which is responsible for the gluon condensate
and confinement, and the scalar field $X(z)$ is introduced to mimic chiral dynamics. Evolution of the dilaton field and scalar
field in 5D resemble the renormalization group from ultraviolet (UV) to infrared (IR). This DhQCD model describes the scalar
glueball spectra and the light meson spectral quite well \cite{Li:2012ay,Li:2013oda,Li:2013xpa}. Further
studies \cite{Li:2014hja,Li:2014dsa,Chelabi:2015cwn} show that this DhQCD model can also describe QCD phase transition,
equation of state of QCD matter and temperature dependent transport properties, including shear viscosity,
bulk viscosity, electric conductivity as well as jet quenching parameter.

For the scalar glueball spectra, it was shown in Ref.\cite{Li:2013oda} that, comparing with the results in the hard-wall
and soft-wall holographic QCD models \cite{glueball-bottomup},  the scalar glueballs including the lowest state and
excited states can be surprisingly well described in the DhQCD model . However, the scalar glueball $0^{++}$ has the
same quantum number as the scalar quarkonium ${\bar q} q$ and tetraquark
${\bar q} q{\bar q} q$ \cite{Jaffe-tetraquark} states, and the complexity of determining the glueball states lies in that gluonic
bound states might always mix with ${\bar q} q$ and ${\bar q} q{\bar q} q$ states. For example, one has to distinguish the
lightest scalar glueball state among 19 scalar mesons observed in the energy range below $2~ {\rm GeV}$ \cite{scalar,mixing}.
Therefore, it is interesting to investigate odd glueballs with unconventional quantum numbers which cannot be carried by
quark-antiquark bound states. These include $J^{PC}=0^{--},0^{+-},2^{+-},3^{-+}$ glueballs, which can only be made
of at-least three-gluon bound states.

This motivate us to investigate the whole glueball spectra including (scalar, vector as well as tensor glueballs and their excitations) in
the framework of the DhQCD model.  The paper is organized as following: In Sec.\ref{sec-glueball-operator} we give the operators of
two-gluon and trigluon glueballs. We introduce the dynamical soft-wall holographic QCD model in Sec.\ref{sec-qdhm}, and calculate the
the glueball spectra in the dynamical holographic QCD model in Sec. \ref{sec-G-Glueball}, it is found that higher-spin glueballs are very
heavy comparing with lattice data, and the even and odd parities cannot be distinguished. Therefore, we introduce a deformed 5-dimension mass
for higher spin glueballs, and in order to distinguish glueballs with even and odd parities, we introduce the positive and negative coupling
between the dilaton field and glueballs.With this set-up, we calculate the glueball spectra in the modified dynamical holographic QCD model
in Sec. \ref{sec-glueball-modifiedm5} and find that the two-gluon
 glueball spectra are in good agreement with lattice data and the trigluon glueball spectra agree with results from QCD sum rules.
 Finally, a short summary is given in Sec.\ref{sec-sum}.

\section{Two-gluon and trigluon glueballs}
\label{sec-glueball-operator}

The AdS/CFT correspondence establishes a one-to-one correspondence between a certain class of $4D$ local
operators  in the ${\cal N}=4$ superconformal gauge theory and $5D$ supergravity fields representing the holographic correspondents
in the $AdS_{5}\times S^{5}$ bulk theory. According to AdS/CFT dictionary, the conformal dimension of a ($f$-form) operator
on the ultraviolet (UV) boundary is related to the $M_5^{2}$ of its dual field in the bulk as follows \cite{Maldacena:1997re,Gubser:1998bc,Witten:1998qj} :
\begin{equation}
M_5^2=(\Delta-f)(\Delta+f-4)\;.   \label{Eq-m5}
\end{equation}

In the bottom-up approach, for example in the holographic QCD models, one can expect a
more general correspondence, i.e. each operator ${\cal O}(x)$ in the 4D field theory corresponds to a field $O(x,z)$ in the 5D bulk theory.
To investigate the glueball spectra, we consider the lowest dimension operators with the corresponding quantum numbers and defined in
the field theory living on the $4D$ boundary. We show the two-gluon and trigluon glueball operators and their corresponding
$5D$ masses in Table \ref{twotrigluon-glueball}.

\begin{table}[!h]
\begin{center}
\begin{tabular}{|c|c|c|c|c|c|}
\hline
$J^{PC}$ & $4D: \mathscr{O}(x)$ & $\Delta$ & $f$ & $M_5^2$ \\
\hline
$0^{++}$ & $Tr(G^2)$ & 4 & 0 & 0 \\
\hline
$0^{--}$ & $Tr(\tilde{G}\{D_{\mu_1}D_{\mu_2}G,G\})$ & 8 & 0  & 32 \\
\hline
$0^{-+}$ & $Tr(G\tilde{G})$ & 4 & 0  & 0 \\
\hline
$1^{\pm-}$ & $Tr(G\{G,G\})$ & 6 & 1 & 15 \\
\hline
$2^{++}$ & $Tr(G_{\mu\alpha}G_{\alpha\nu}-\frac{1}{4}\delta_{\mu\nu}G^{2})$ & 4 & 2 & 4 \\
\hline
$2^{++}$ & $E_{i}^{a}E_{j}^{a}-B_{i}^{a}B_{j}^{a}-trace$ & 4 & 2 & 4 \\
\hline
$2^{-+}$ & $E_{i}^{a}B_{j}^{a}+B_{i}^{a}E_{j}^{a}-trace$ & 4 & 2  & 4 \\
\hline
$2^{\pm-}$ & $Tr(G\{G,G\})$ & 6 & 2 & 16 \\
\hline
\end{tabular}
\caption{5D mass square of two-gluon and trigluon glueballs.The operators are taken from \cite{glueball-AdS-operator} and
\cite{Qiao:2014,Qiao:2015iea}.}
\label{twotrigluon-glueball}
\end{center}
\end{table}

For trigluon glueball $0^{--}$,  the detailed structure of the operator is given in Ref. \cite{Qiao:2014}
\begin{eqnarray}
j^A_{0^{--}}&\sim & d^{a b c} [g^t_{\alpha \beta} \tilde{G}^a_{\mu \nu}][\partial_\alpha \partial_\beta G^b_{\nu \rho}][G^c_{\rho \mu}],\\
j^B_{0^{--}}&\sim & d^{a b c} [g^t_{\alpha \beta} G^a_{\mu \nu}][\partial_\alpha \partial_\beta \tilde{G}^b_{\nu \rho}][G^c_{\rho \mu}],\\
j^C_{0^{--}}&\sim & d^{a b c} [g^t_{\alpha \beta} G^a_{\mu \nu}][\partial_\alpha \partial_\beta G^b_{\nu \rho}][\tilde{G}^c_{\rho \mu}],\\
j^D_{0^{--}}&\sim & d^{a b c} [g^t_{\alpha \beta} \tilde{G}^a_{\mu \nu}][\partial_\alpha \partial_\beta \tilde{G}^b_{\nu \rho}][\tilde{G}^c_{\rho \mu}],
\end{eqnarray}
where $d^{abc}$ stands for the totally symmetric $SU_c(3)$ structure constant and $g^t_{\alpha \beta}= g_{\alpha \beta}- \partial_\alpha \partial_\beta/\partial^2$.

The interpolating currents of the $2^{+-}$ oddball $2^{+-}$ takes the form as \cite{Qiao:2015iea},
\begin{eqnarray}
j^{2^{+-}, \; A}_{\mu \alpha}(x) & \!\! = \!\! & g_s^3 d^{a b c} [G^a_{\mu \nu}(x)][G^b_{\nu \rho}(x)][G^c_{\rho \alpha}(x)]\, ,\label{current-2+-A}\\
j^{2^{+-}, \; B}_{\mu \alpha}(x) & \!\! = \!\! & g_s^3 d^{a b c} [G^a_{\mu \nu}(x)][\tilde{G}^b_{\nu \rho}(x)][\tilde{G}^c_{\rho \alpha}(x)]\, ,  \label{current-2+-B}\\
j^{2^{+-}, \; C}_{\mu \alpha}(x) & \!\! = \!\! & g_s^3 d^{a b c} [\tilde{G}^a_{\mu \nu}(x)][G^b_{\nu \rho}(x)][\tilde{G}^c_{\rho \alpha}(x)]\, ,  \label{current-2+-C}\\
j^{2^{+-}, \; D}_{\mu \alpha}(x) & \!\! = \!\! & g_s^3 d^{a b c} [\tilde{G}^a_{\mu \nu}(x)][\tilde{G}^b_{\nu \rho}(x)][G^c_{\rho \alpha}(x)]\, .\label{current-2+-D}
\end{eqnarray}

%\section{Glueball spectra in the dynamical soft-wall holographic QCD model}
%\label{sec-glueball-dhQCD}

\section{The dynamical soft-wall holographic QCD model and gluodynamics}
\label{sec-qdhm}

The dynamical soft-wall holographic QCD model is described in Ref.\cite{Li:2013oda}.
The pure gluon part of QCD can be modelled by the 5D graviton-dilaton coupled action:
\begin{eqnarray}\label{action-graviton-dilaton}
 S_G=\frac{1}{16\pi G_5}\int
 d^5x\sqrt{g_s}e^{-2\Phi}\left(R_s+4\partial_M\Phi\partial^M\Phi-V^s_G(\Phi)\right),
\end{eqnarray}
where $G_5$ is the 5D Newton constant, $g_s$, $\Phi$ and $V_G^s$ are the 5D
metric, the dilaton field and dilaton potential in the string frame, respectively.
The metric is chosen to be
\begin{eqnarray}\label{metric-ansatz}
g^s_{MN}=b_s^2(z)(dz^2+\eta_{\mu\nu}dx^\mu dx^\nu), ~ ~ b_s(z)\equiv e^{A_s(z)}.
\end{eqnarray}

Under the conformal transformation
\begin{equation}
g^E_{MN}=g^s_{MN}e^{-4\Phi/3}, ~~ V^E_G=e^{4\Phi/3}V_{G}^s,
\end{equation}
Eq.(\ref{action-graviton-dilaton}) can be rewriten in the Einstein frame
\begin{eqnarray}\label{graviton-dilaton-E}
S_G^E=\frac{1}{16\pi G_5}\int d^5x\sqrt{g_E}\left(R_E-\frac{4}{3}\partial_M\Phi\partial^M\Phi-V_G^E(\Phi)\right).
\end{eqnarray}
The Einstein equations are
\begin{eqnarray}
E_{MN}+\frac{1}{2}g^E_{MN}\left(\frac{4}{3}\partial_L\Phi\partial^L\Phi+V_G^E(\Phi)\right)
-\frac{4}{3}\partial_M\Phi\partial_N\Phi&=&0,\\
 \frac{8}{3\sqrt{g_E}}\partial_M(\sqrt{g_E}\partial^M\Phi)-\partial_{\Phi}V_G^E(\Phi)&=&0.
\end{eqnarray}
Substituting the metric of Eq.(\ref{metric-ansatz}) into the above equations, we can obtain:
\begin{eqnarray}
-A_E^{''}+A_E^{'2}-\frac{4}{9}\Phi^{'2}=0, \\
\label{AEPhi}
\Phi^{''}+3A_E^{'}\Phi^{'}-\frac{3}{8}e^{2A_E}\partial_\Phi V_G^E(\Phi)=0,
\label{AEVPhi2}
\end{eqnarray}
where
\begin{equation}
b_E(z)=b_s(z)e^{-\frac{2}{3}\Phi(z)}=e^{A_E(z)},~~A_E(z)=A_s(z)-\frac{2}{3}\Phi(z).
\end{equation}
In the string frame, the above two equations of motion are
\begin{eqnarray}
-A_s^{''}-\frac{4}{3}\Phi^{'}A_s^{'}+A_s^{'2}+\frac{2}{3}\Phi^{''}=0, \\
\Phi^{''}+(3A_s^{'}-2\Phi^{'})\Phi^{'}-\frac{3}{8}e^{2A_s-\frac{4}{3}\Phi}\partial_\Phi (e^{\frac{4}{3}\Phi}V_G^s(\Phi))=0.
\end{eqnarray}

We take the same dilaton field as that in the KKSS model or soft-wall holographic QCD model \cite{Karch:2006pv}, i.e.,
\begin{equation}
\Phi=\mu_G^2 z^2.
\label{dilaton}
\end{equation}
It is simple to solve the metric $A_E$ and the dilaton potential $V_G^E(\Phi)$ in the quadratic dilaton background
\begin{eqnarray}\label{puregluosol}
 A_E(z) & = &\log(\frac{L}{z})-\log(_0F_1(5/4,\frac{\Phi^2}{9})), \\
 V_G^E(\Phi)& = &-\frac{12 _0F_1(1/4,\frac{\Phi^2}{9})^2}{L^2}
 +\frac{16 _0F_1(5/4,\frac{\Phi^2}{9})^2\Phi^2}{3L^2},
\end{eqnarray}
with $_0F_1(a;z)$ the hypergeometric function.

\section{Glueball spectra in the dynamical soft-wall holographic QCD model}
\label{sec-G-Glueball}

\subsection{Scalar glueballs}
The 5D action for the scalar glueball $\mathscr{G}(x,z)$ in the string frame takes the form as that in the original soft-wall model \cite{Colangelo:2007pt,Forkel:2007ru}
\begin{eqnarray}
S_{\mathscr{G}}=-\int d^5 x \sqrt{g_s}\frac{1}{2}e^{-\Phi}\big[ \partial_M \mathscr{G}\partial^M
\mathscr{G}+M_{\mathscr{G},5}^2 \mathscr{G}^2\big].
\end{eqnarray}
It is notice that the metric structure in the dynamical soft-wall model is solved from Eq. (\ref{AEPhi}) instead of ${\rm AdS}_5$.

The Equation of motion for the scalar glueballs $\mathscr{G}$ is given below as
\begin{eqnarray}
-e^{-(3A_s-\Phi)}\partial_z(e^{3A_s-\Phi}\partial_z\mathscr{G}_n)+e^{2A_s}M_{\mathscr{G},5}^2\mathscr{G}_n=m_{\mathscr{G},n}^2 \mathscr{G}_n.
\end{eqnarray}
Via the substitution $\mathscr{G}_n \rightarrow e^{-\frac{1}{2}(3A_s-\Phi)}\mathscr{G}_n$, the equation can be brought into schr\"{o}dinger-like equation
\begin{eqnarray}
-\mathscr{G}_n^{''}+V_{\mathscr{G}} \mathscr{G}_n=m_{\mathscr{G},n}^2 \mathscr{G}_n,
\label{EOM-glueball}
\end{eqnarray}
with the 5D effective schr\"{o}dinger potential
\begin{equation}
V_{\mathscr{G}}=\frac{3A_s^{''}-\Phi^{''}}{2}+\frac{(3A_s^{'}-\Phi^{'})^2}{4}+e^{2A_s}M_{\mathscr{G},5}^2.
\label{potential-glueball-s}
\end{equation}

\subsection{Vector glueballs}

For vector gluebalsl $\mathscr{V}$, the 5D action is
\begin{eqnarray}
S_{V}=-\int d^{5}x\sqrt{g}e^{-\Phi}(\frac{1}{4}F^{MN}F_{MN}+\frac{1}{2}M_{\mathscr{V},5}^2\mathscr{V}^{2}),
\end{eqnarray}
where $F_{MN}=\partial_M\mathscr{V}_N-\partial_N\mathscr{V}_M$.

The equation of motion for vector glueballs $\mathscr{V}$ is
\begin{eqnarray}
-e^{-(A_s-\Phi)}\partial_z(e^{A_s-\Phi}\partial_z\mathscr{V}_n)+e^{2A_s}M_{\mathscr{V},5}^2\mathscr{V}_n=m_{\mathscr{V},n}^2 \mathscr{V}_n.
\end{eqnarray}
Via the substitution $\mathscr{V}_n \rightarrow e^{-\frac{1}{2}(A_s-\Phi)}\mathscr{V}_n$, the equation can be brought into schr\"{o}dinger-like equation
\begin{eqnarray}
-\mathscr{V}_n^{''}+V_{\mathscr{V}} \mathscr{V}_n=m_{\mathscr{V},n}^2 \mathscr{V}_n,
\end{eqnarray}
with the 5D effective schr\"{o}dinger potential
\begin{equation}
V_{\mathscr{V}}=\frac{A_s^{''}-\Phi^{''}}{2}+\frac{(A_s^{'}-\Phi^{'})^2}{4}+e^{2A_s}M_{\mathscr{V},5}^2.
\label{potential-glueball-v}
\end{equation}

\subsection{Tensor glueballs}

For tensor glueballs, the 5D action is
\begin{eqnarray}
S_{T}=-\frac{1}{2}\int d^{5}x\sqrt{g}e^{-\Phi}(&&\nabla_{L}h_{MN}\nabla^{L}h^{MN}-2\nabla_{L}h^{LM}\nabla^{N}h_{NM}+2\nabla_{M}h^{MN}\nabla_{N}h\nonumber\\
&&-\nabla_{M}h\nabla^{M}h+M_{h,5}^{2}(h^{MN}h_{MN}-h^{2})),
\end{eqnarray}
where $h=g^{MN}h_{MN}$. With the constraint
\begin{eqnarray}
\nabla_Mh^{MN}=0,~~h=0,~~h_{\mu\nu}=e^{2A_s}\mathscr{H}_{\mu\nu},~~h_{Mz}=0,
\end{eqnarray}

The equation of motion for tensor glueballs $\mathscr{H}_{\mu\nu}$ is
\begin{eqnarray}
-e^{-(3A_s-\Phi)}\partial_z(e^{3A_s-\Phi}\partial_z\mathscr{H}_n)+e^{2A_s}M_{\mathscr{H},5}^2\mathscr{H}_n=m_{\mathscr{H},n}^2\mathscr{H}_n.
\end{eqnarray}
Via the substitution $\mathscr{H}_n \rightarrow e^{-\frac{1}{2}(3A_s-\Phi)}\mathscr{H}_n$, the equation can be brought into schr\"{o}dinger-like equation
\begin{eqnarray}
-\mathscr{H}_n^{''}+V_{\mathscr{H}} \mathscr{H}_n=m_{\mathscr{H},n}^2 \mathscr{H}_n,
\end{eqnarray}
with the 5D effective schr\"{o}dinger potential
\begin{equation}
V_{\mathscr{H}}=\frac{3A_s^{''}-\Phi^{''}}{2}+\frac{(3A_s^{'}-\Phi^{'})^2}{4}+e^{2A_s}M_{\mathscr{H},5}^2.
\label{potential-glueball-t}
\end{equation}

\subsection{Numerical results}
\label{sec-glueball-hwsw}

%\subsection{Glueball spectra in the hard-wall holographic QCD model}
For numerical calculations, we have to fix parameters in the model. In the dynamical holographic model, there is only one free
parameter, i.e., $\mu_G$. We fix this parameter by fitting the scalar glueballs spectra from lattice results \cite{Meyer:2004gx,Lucini:2001ej,Morningstar:1999rf,Chen:2005mg} as shown in Table \ref{Lat-glueballspectra}.
The lattice data in Table \ref{Lat-glueballspectra} indicates the slope of the Regge spectra is around $4{\rm GeV}^2$,
which is equivalent to $\mu_G\simeq 1 {\rm GeV}$ in the dynamical holographic QCD model.

\begin{table}[!h]
\begin{center}
\begin{tabular}{cccccccc}
\hline\hline
n($0^{++}$) & ~Lat1   &   Lat2 &   Lat3  &   Lat4 &    Lat5            \\
\hline
  ~&  $N_c=3$  &$N_c=3$  & $N_c\rightarrow \infty $  &$N_c=3$  &$N_c=3$  \\
   1 & $1475(30)(65)$     & 1580(11)   &   1480(07)   &1730(50)(80)  &1710(50)(80)\\
   2 & $2755(70)(120)$   & 2750(35)   &   2830(22)  &2670(180)(130)   &   \\
   3 & $3370(100)(150)$      & ~        &        &       &      \\
   4 & $3990(210)(180)$      & ~        &        &       &      \\
\hline
\end{tabular}
\caption{Lattice data for $0^{++} glueball$ in unit of ${\rm MeV}$. Lat1 data
from Ref.\cite{Meyer:2004gx}, Lat2 and Lat3 data from Ref.\cite{Lucini:2001ej},
Lat4 \cite{Morningstar:1999rf} and Lat5 \cite{Chen:2005mg} are anisotropic results.}
\label{Lat-glueballspectra}
\end{center}
\end{table}

We will also compare our results in the dynamical holographic QCD model with those in the hard-wall
and soft-wall holographic QCD models. In the hard-wall holographic QCD model, the equation of motion for glueball $\mathscr{G}$ is:
\begin{eqnarray}
-e^{-cA_s}\partial_z(e^{cA_s}\partial_z\mathscr{G}_n)+e^{2A_s}M_{\mathscr{G},5}^2\mathscr{G}_n=m_{\mathscr{G},n}^2 \mathscr{G}_n,
\end{eqnarray}
where $c=1$ for vector glueballs and $c=3$ for scalar and tensor glueballs. With UV boundary condition $\mathscr{G}_n(\epsilon)=0$,
the solution is:
\begin{eqnarray}
\mathscr{G}_n(z)=z^{\frac{1+c}{2}}J_n(m_{\mathscr{G},n}, z),
\end{eqnarray}
where $n=\sqrt{1+2c+c^2+4M_{\mathscr{G},5}^2}/2$ and $J$ is Bessel function.
IR boundary condition $\partial_z\mathscr{G}_n(z_m)=0$ gives the discrete spectrum of the glueballs. Here $z_m$
is the hard cut-off, which is the only parameter in the hard-wall model, and can be fixed by the ground state of the scalar
glueball $0^{++}$. When we take the mass for the lowest scalar glueball as $1730 {\rm MeV}$, which fixes $z_m=452{\rm MeV}$
in the hard wall model.

In the soft-wall model, with the dilaton background takes the quadratic form $\Phi=\mu_G^2z^2$ and the metric structure is
still ${\rm AdS}_5$, the Regge spectra for glueball can be derived as \cite{Colangelo:2007pt,Forkel:2007ru}
\begin{equation}
m_{n}^{2}=\mu_G^2\left \{4n+c+1+\sqrt{(c+1)^{2}+4M_{5}^{2}} \right \}, ~ n=0,1,2, \cdots
\label{glueballmass-sw}
\end{equation}
where $c=3$ in case of scalar and tensor and $c=1$ in case of vector. There is also only one parameter $\mu_G$ in the soft-wall model.
The Regge slope of $0^{++}$ glueball gives $\mu_G=1 {\rm GeV}$ (SW$\dagger$), and the the lowest scalar glueball mass $1730 {\rm MeV}$
gives $\mu_G=0.5 {\rm GeV}$ (SW$\dagger\dagger$).

\begin{table}[!h]
\begin{center}
\begin{tabular}{|c|c|c|c|c|c|c|}
\hline
$J^{PC}$ & Lattice & HW & SW$\dagger$ & SW$\ddagger$ & DSW  \\
\hline
$0^{++}$ & 1475-1730 & 1730 & 1730 & 2828 & 1593  \\
\hline
$0^{*++}$ & 2670-2830 & 3168 & 2119 & 3464 & 2618  \\
\hline
$0^{**++}$ & 3370  & 4593 & 2447 & 4000 & 3311  \\
\hline
$0^{***++}$ & 3990 & 6016 & 2735 & 4472 & 3877 \\
\hline
\end{tabular}
\caption{The mass spectra of $0^{++}$ glueballs in the dynamical soft-wall model, compared with
 combined lattice data \cite{Meyer:2004gx,Lucini:2001ej,Morningstar:1999rf,Chen:2005mg}, and
 hard-wall model ($z_m=452{\rm MeV}$), soft-wall model with SW$\dagger$ indicates $\mu_G$
 is fixed by lowest mass of $0^{++}$ glueball, and SW$\ddagger$ indicates $\mu_G$ is fixed by Regge
 slope of $0^{++}$ glueball. The unit is in {\rm MeV}.}
\label{Table-scalar-glueball}
\end{center}
\end{table}

Table \ref{Table-scalar-glueball} shows scalar glueball spectra in the dynamical soft-wall holographic QCD model,
and the results are compared with combined lattice data  \cite{Meyer:2004gx,Lucini:2001ej,Morningstar:1999rf,Chen:2005mg}
as well as hard-wall and soft-wall models, respectively. It is found that the hard-wall model cannot produce Regge spectra,
and the soft-wall  model cannot simultaneously produce the correct Regge slope and the ground state of the scalar glueball.
As it was shown in Ref. \cite{Li:2013oda},  the Regge slope and the ground state of the scalar glueball
can be correctly produced in the dynamical soft-wall holographic QCD model with only one parameter.

\begin{table}[!h]
\begin{center}
\begin{tabular}{|c|c|c|c|c|c|c|c|}
\hline
$J^{PC}$ & Lattice & HW1  & SW$\dagger$ & SW$\ddagger$ & DSW  \\
\hline
$0^{-+}$ & 2590 & 1730  & 1730 & 2828 & 1593  \\
\hline
$0^{*-+}$ & 3640 & 3168  & 2119 & 3464 & 2618  \\
\hline
$0^{--}$ & 5166 & 3658 & 2447 & 4000 & 10759  \\
\hline
$1^{+-}$ & 2940 & 2571  & 1934 & 3162 & 7535 \\
\hline
$1^{--}$ & 3850 & 2571  & 1934 & 3162 & 7535  \\
\hline
$2^{++}$ & 2400 & 2134  & 1901 & 3108 & 4328  \\
\hline
$2^{-+}$ & 3100 & 2134 & 1901 & 3108 & 4328  \\
\hline
$2^{*-+}$ & 3890 & 3646 & 2260 & 3696 & 5233  \\
\hline
$2^{+-}$ & 4140 & 2927  & 2201 & 3598 & 7830  \\
\hline
$2^{--}$ & 3930 & 2927  & 2201 & 3598 & 7830  \\
\hline
\end{tabular}
\caption{The mass spectra of vector and tensor glueballs in the dynamical soft-wall model, compared with
lattice data, and hard-wall model ($z_m=452{\rm MeV}$), soft-wall model with SW$\dagger$ indicates $\mu_G$
 is fixed by lowest mass of $0^{++}$ glueball, and SW$\ddagger$ indicates $\mu_G$ is fixed by Regge
slope of $0^{++}$ glueball. The unit is in {\rm MeV}.}
\label{Table-VT-glueballall}
\end{center}
\end{table}

With the parameter fixed by the scalar glueball, we calculate the vector and tensor glueball masses in the dynamical
soft-wall model and compare with lattice data as well as results from hard-wall and soft-wall models.
The results are shown in Table \ref{Table-VT-glueballall}. It is observed that for vector and tensor glueballs, the results
from the dynamical holographic QCD model are far away from lattice data, especially the masses for higher spin states
are too heavy comparing with lattice data. The glueball massea from the hard-wall model are in general lighter than lattice
results. Among the three models, the soft-wall model (SW$\dagger\dagger$) with the parameter fixed by the Regge slope
can produce reasonable good results comparing with lattice data. However, all models cannot distinguish even and odd parity
state for the glueball  with the same spin.

In the next section, we will improve the dynamical soft-wall  holographic QCD model in order to produce reasonable
glueball spectra.

%\subsection{Glueball spectra in the soft-wall holographic QCD model}

\section{Glueball spectra in modified dynamical soft-wall holographic QCD model}
\label{sec-glueball-modifiedm5}

As we observed from last section that the masses for higher spin glueballs are too heavy comparing with lattice data,
while these states are reasonable in the soft-wall model. This indicates that only scalar glueballs are sensitive to the
deformed metric, and other glueballs are not excited from this deformed metric background. Therefore we introduce
a deformed 5D mass squared for glueballs. In order to distinguish even and odd
parity, we introduce the positive and negative coupling between the dilaton field and glueballs, respectively. With this set-up,
now the 5D action for the scalar, vector and tensor glueballs $\mathscr{G}(x,z)$ take the following form:
\begin{eqnarray}
S_{\mathscr{G}}=-\frac{1}{2}\int d^5 x \sqrt{g_s}e^{-p\Phi}(&&\partial_M \mathscr{G}\partial^M\mathscr{G}+M_{\mathscr{G},5}^2(z) \mathscr{G}^2), \\
S_{V}=-\frac{1}{2}\int d^{5}x\sqrt{g_s}e^{-p\Phi}(&&\frac{1}{2}F^{MN}F_{MN}+M_{\mathscr{V},5}^2(z) \mathscr{V}^{2}), \\
S_{T}=-\frac{1}{2}\int d^{5}x\sqrt{g_s}e^{-p\Phi}(&&\nabla_{L}h_{MN}\nabla^{L}h^{MN}-2\nabla_{L}h^{LM}\nabla^{N}h_{NM}+2\nabla_{M}h^{MN}\nabla_{N}h\nonumber\\
&&-\nabla_{M}h\nabla^{M}h+M_{h,5}^{2}(z)(h^{MN}h_{MN}-h^{2})),
\end{eqnarray}
where $M_5^2(z)=M_5^2e^{-2\Phi /3}$, $p=1$ for even parity and $p=-1$ for odd parity.

The equation of motion for any glueball $\mathscr{A}$ can be brought into schr\"{o}dinger-like equation
\begin{eqnarray}
-\mathscr{A}_n^{''}+V_{\mathscr{A}} \mathscr{A}_n=m_{\mathscr{A},n}^2 \mathscr{A}_n,
\end{eqnarray}
with the 5D effective schr\"{o}dinger potential
\begin{equation}
V_{\mathscr{A}}=\frac{c A_s^{''}-p\Phi^{''}}{2}+\frac{(c A_s^{'}-p\Phi^{'})^2}{4}+e^{2A_s-\frac{2}{3}\Phi}M_{\mathscr{A},5}^2,
\label{potential-glueballmodified}
\end{equation}
where $c=1$ for 1-form and $c=3$ for 0-form and 2-form, and $M_{\mathscr{A},5}^2$ is the value given in Table \ref{twotrigluon-glueball}.

\begin{figure}[h]
\begin{center}
\epsfxsize=5.5 cm \epsfysize=5.5 cm \epsfbox{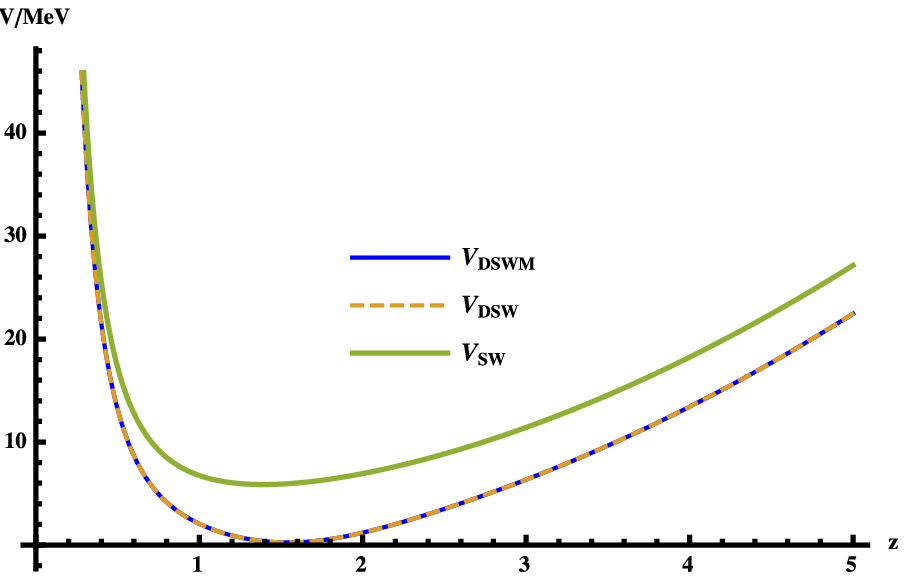} \hspace*{0.2cm}
\hskip 0.15 cm
\textbf{( $0^{++}$ ) } \\
\epsfxsize=5.5 cm \epsfysize=5.5 cm \epsfbox{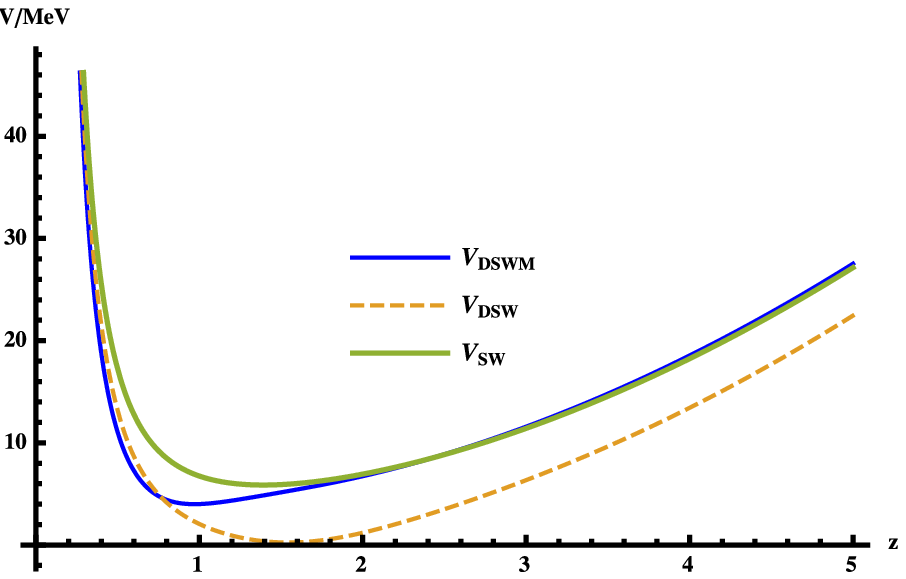} \hspace*{0.2cm}
\epsfxsize=5.5 cm \epsfysize=5.5 cm \epsfbox{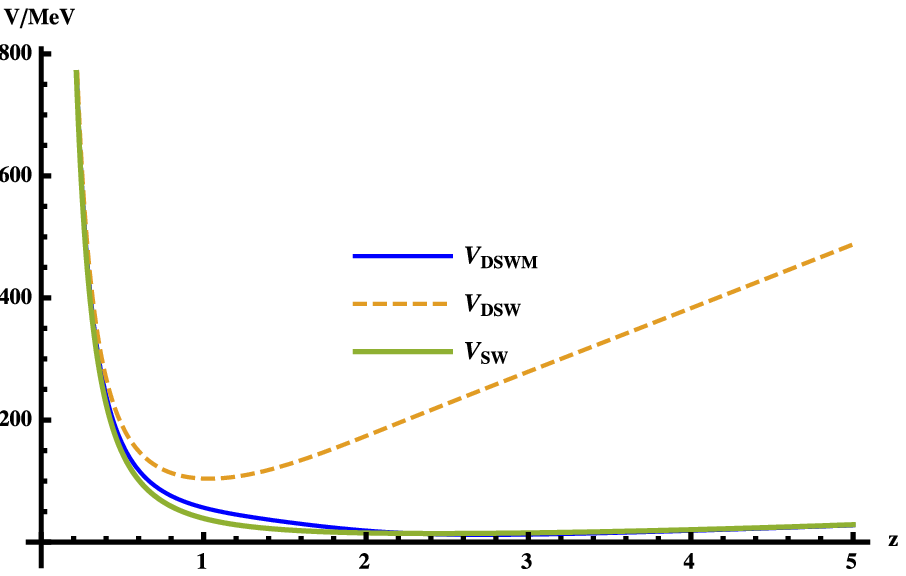} \vskip -0.005cm
\hskip 0.15 cm
\textbf{( $0^{-+}$ ) } \hskip 6.5 cm \textbf{( $0^{--}$ )} \\
\epsfxsize=5.5 cm \epsfysize=5.5 cm \epsfbox{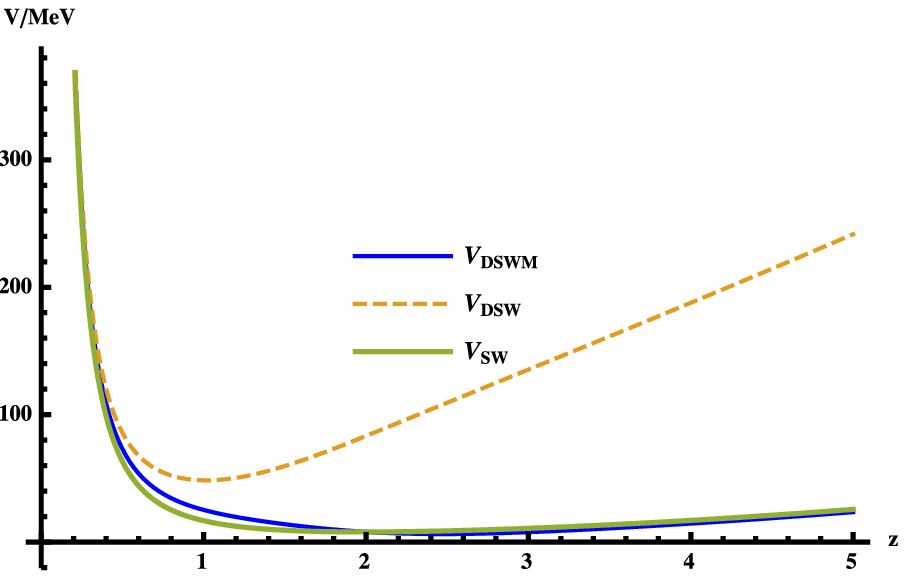} \hspace*{0.2cm}
\epsfxsize=5.5 cm \epsfysize=5.5 cm \epsfbox{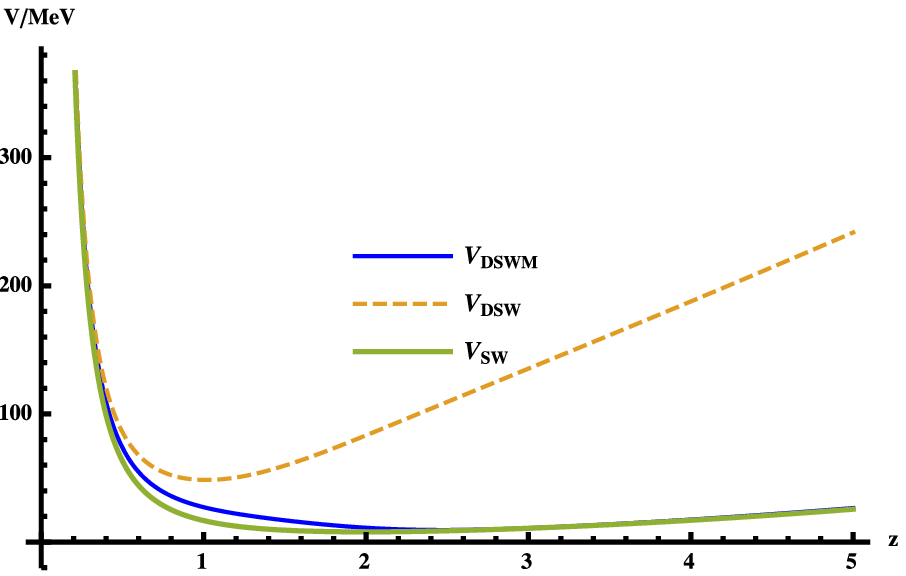} \vskip -0.005cm
\hskip 0.15 cm
\textbf{( $1^{+-}$ ) } \hskip 6.5 cm \textbf{( $1^{--}$ )} \\
\end{center}
\caption{The effective schr\"{o}dinger potential $V$ of scalar and vector glueballs in the soft-wall model (green thick line),
dynamical soft-wall model (orange dashed line) and dynamical soft-wall model with modified $M_5^2$(blue line). }
\label{vg1}
\end{figure}

\begin{figure}[h]
\begin{center}
\epsfxsize=5.5 cm \epsfysize=5.5 cm \epsfbox{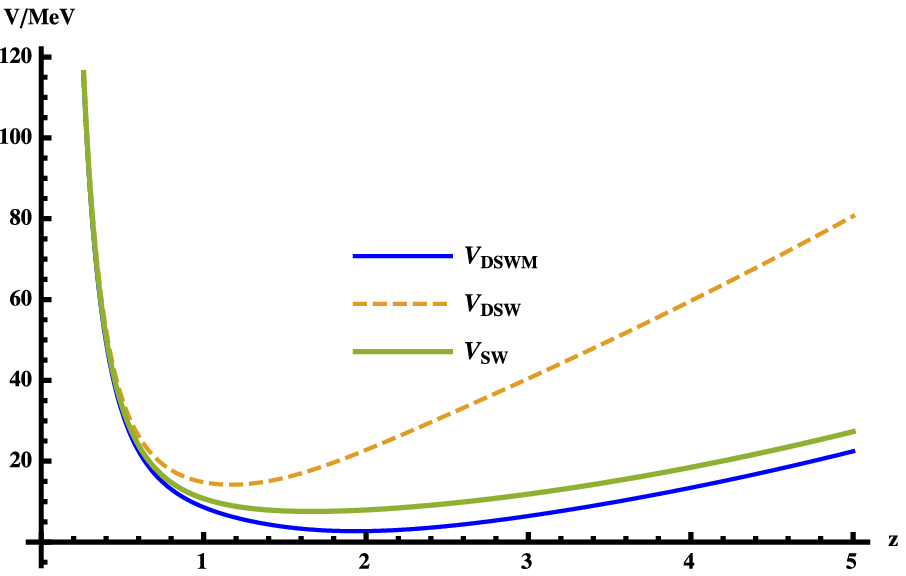} \hspace*{0.2cm}
\epsfxsize=5.5 cm \epsfysize=5.5 cm \epsfbox{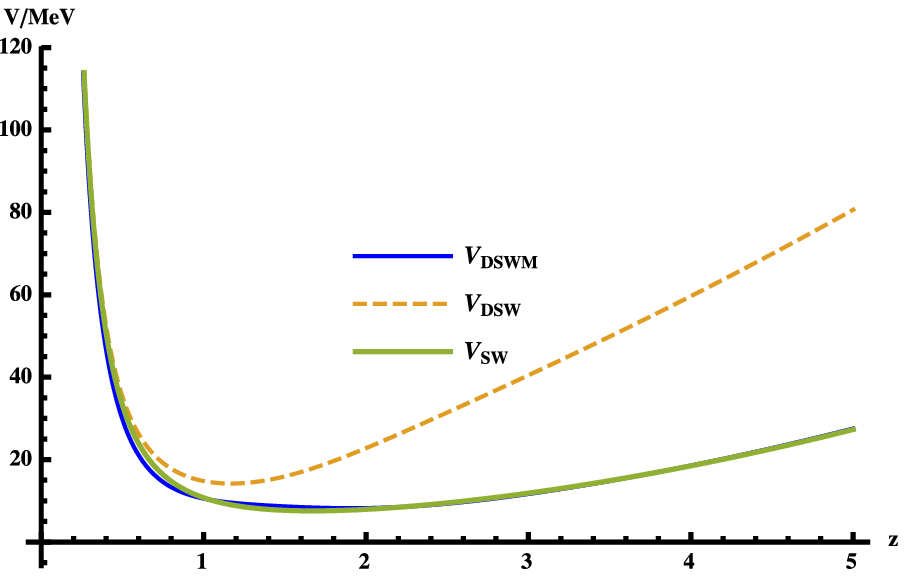} \vskip -0.005cm
\hskip 0.15 cm
\textbf{( $2^{++}$ ) } \hskip 6.5 cm \textbf{( $2^{-+}$ )} \\
\epsfxsize=5.5 cm \epsfysize=5.5 cm \epsfbox{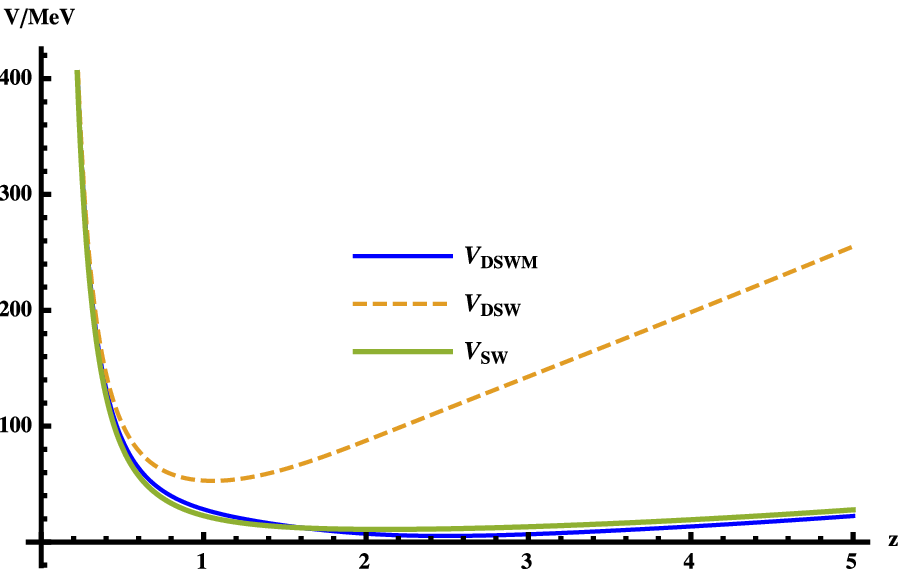} \hspace*{0.2cm}
\epsfxsize=5.5 cm \epsfysize=5.5 cm \epsfbox{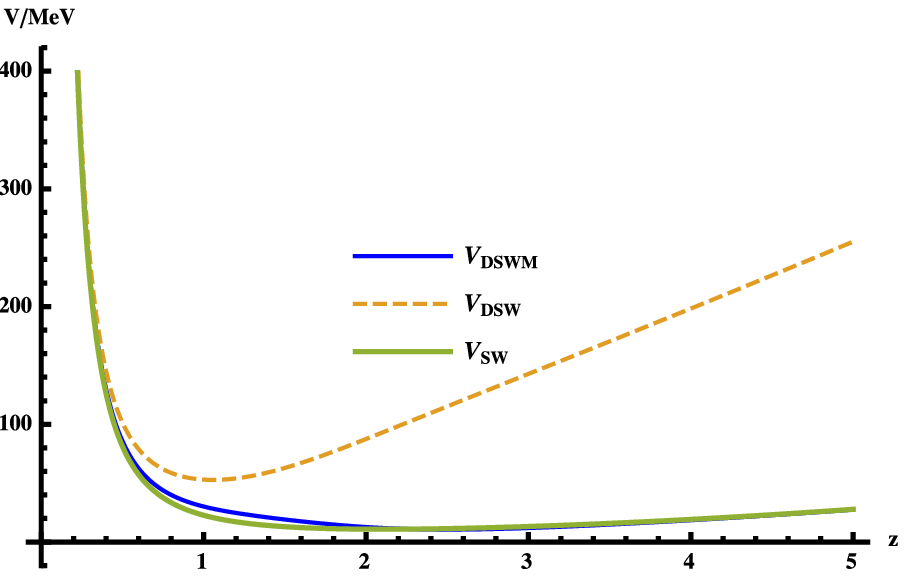} \vskip -0.005cm
\hskip 0.15 cm
\textbf{( $2^{+-}$ ) } \hskip 6.5 cm \textbf{( $2^{--}$ )} \\
\end{center}
\caption{The effective schr\"{o}dinger potential $V$ of tensor glueballs in the soft-wall model (green thick line),
dynamical soft-wall model (orange dashed line) and dynamical soft-wall model with modified $M_5^2$(blue line). }
\label{vg2}
\end{figure}

Compare the 5D effective schr\"{o}dinger potential Eq.(\ref{potential-glueballmodified}) with Eqs.(\ref{potential-glueball-s}),
(\ref{potential-glueball-v}) and (\ref{potential-glueball-t}), we can see the effect of the
deformed 5D mass square $M_5^2(z)=M_5^2e^{-2\Phi /3}$ is to counteract the deform metric background. In Fig. \ref{vg1}
and Fig. \ref{vg2}, we show the 5D effective schr\"{o}dinger potential Eq.(\ref{potential-glueballmodified}) as a function
of $z$ and compare with results from the soft-wall model and the original dynamical soft-wall model. It is found that at infrared (IR),
except the scalar glueball $0^{++}$, the 5D effective schr\"{o}dinger potential for other glueballs
in the modified dynamical soft-wall holographic QCD model coincide with those from soft-wall model. The parity
difference $p=\pm$ only brings the difference of the 5D effective schr\"{o}dinger potential in the range of $0.5<z<2$.

\begin{figure}[!htb]
\begin{center}
\includegraphics[width=0.9\columnwidth]{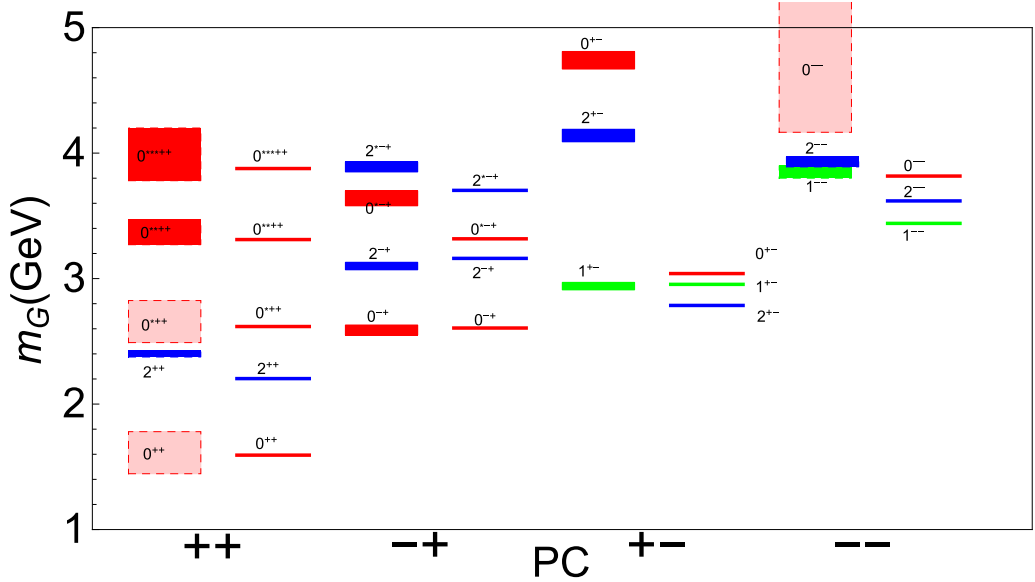}
\caption{The mass spectra of glueballs in the modified dynamical soft-wall model with $\mu=1~{\rm GeV}$ (line) compared with the lattice result \cite{Meyer:2004gx,Lucini:2001ej,Morningstar:1999rf,Chen:2005mg,Gregory:2012hu} (rectangle).}
\label{gb}
\end{center}
\end{figure}

The final results of the glueball spectra in the modified dynamical holographic QCD model are shown in Fig. \ref{gb} and  in Table \ref{odd-glueballspectra} with details. It is found that with only one parameter $\mu_G=1{\rm GeV}$, which is fixed by the Regge 
slope of the scalar gluball spectra, one can produce other glueballs spectra agree well with lattice data, except three trigluon glueball 
states $0^{--}$, $0^{+-}$ and $2^{+-}$, whose masses are 1.5 GeV lighter than lattice results. Considering that we
only take the simplest quadratic dilaton profile, which corresponds to dimension-2 gluon condensate (or effectively two-gluon condensate)
in the vacuum, our results might indicate that these three trigluon glueballs $0^{--}$, $0^{+-}$ and $2^{+-}$ are dominated by three-gluon
condensate contribution. 

\begin{table}[!h]
\begin{center}
\begin{tabular}{|c|c|c|c|c|}
\hline\hline
$J^{PC}$ & LQCD & Flux tube model & QCDSR & MDSM \\
\hline
$0^{++}$ & 1.475-1.73 & 1.52 & 1.5 & 1.593 \\
\hline
$0^{*++}$ & 2.67-2.83 & 2.75 & -- & 2.618 \\
\hline
$0^{**++}$ & 3.37 & -- & -- & 3.311 \\
\hline
$0^{***++}$ & 3.99 & -- & -- & 3.877 \\
\hline
$0^{-+}$ & 2.59 & 2.79 & 2.05 & 2.606 \\
\hline
$0^{*-+}$ & 3.64 & -- & -- & 3.317 \\
\hline
$0^{--}$ & 5.166 & 2.79 & 3.81 & 3.817 \\
\hline
$0^{+-}$ & 4.74 & 2.79 & 4.57 & 3.04 \\
\hline
$0^{++}\mathsection$ & -- & -- & 3.1 & 2.667 \\
\hline
$1^{+-}$ & 2.94 & 2.25 & -- & 2.954 \\
\hline
$1^{--}$ & 3.85 & -- & -- & 3.44 \\
\hline
$2^{++}$ & 2.4 & 2.84 & 2 & 2.203 \\
\hline
$2^{-+}$ & 3.1 & 2.84 & -- & 3.161 \\
\hline
$2^{*-+}$ & 3.89 & -- & -- & 3.703 \\
\hline
$2^{+-}$ & 4.14 & 2.84 & 6.06 & 2.786 \\
\hline
$2^{--}$ & 3.93 & 2.84 & -- & 3.619 \\
\hline
\hline
\end{tabular}
\caption{The mass of glueball spectra in Lattice QCD \cite{Meyer:2004gx,Lucini:2001ej,Morningstar:1999rf,Chen:2005mg,Gregory:2012hu}, Flux tube model \cite{Isgur:1984bm}, QCDSR \cite{Narison:1988ts,Latorre:1987wt,Narison:1996fm,scalar,Qiao:2014,Qiao:2015iea} and modified dynamical soft-wall model. Note that $0^{++}\mathsection$ is trigluonium. The unit is in {\rm GeV}.}
\label{odd-glueballspectra}
\end{center}
\end{table}

\section{Conclusion and discussion}
\label{sec-sum}

In this work, we study scalar, vector and tensor glueball spectra in the framework of 5-dimension dynamical holographic
QCD model, where the metric structure is deformed self-consistently by the dilaton field. It is found that only scalar glueballs
are excited from this deformed metric background, and other glueballs excited from this deformed metric background are much
heavy comparing with lattice data. Therefore, for higher spin glueballs, we introduce a deformed 5-dimension mass in order to
counteract the effect of the deformed metric background.  In order to distinguish glueballs with even and odd parities, we introduce the
positive and negative coupling between the dilaton field and glueballs.

With these set-ups, we calculate the glueball spectra with only one free parameter in the dynamical holographic QCD model, 
which is fixed by the scalar glueball spectra.  It is found that all two-gluon glueball spectra produced in the dynamical holographic 
QCD model are in good agreement with lattice data. We investigate six trigluon glueballs, among these trigluon glueballs, the 
produced masses for $1^{\pm -}$ and $2^{--}$ are in good agreement with lattice data, and the
produced masses for $0^{--}$, $0^{+-}$ and $2^{+-}$ are around 1.5 GeV lighter than lattice results. Considering that we
only take the simplest quadratic dilaton profile, which corresponds to dimension-2 gluon condensate (or effectively two-gluon condensate)
in the vacuum, our results might indicate that the three trigluon glueballs $0^{--}$, $0^{+-}$ and $2^{+-}$ are dominated by three-gluon
condensate contribution. Further studies with more complicated dilaton profile are needed.

\vskip 0.5cm
{\bf Acknowledgement}
\vskip 0.2cm

We thank Danning Li,  Hao Ouyang, Cong-Feng Qiao and Liang Tang for useful discussions. This work is supported by 
the NSFC under Grant Nos. 11175251 and 11275213, DFG and NSFC (CRC 110),
CAS key project KJCX2-EW-N01, K.C.Wong Education Foundation, and
Youth Innovation Promotion Association of CAS.

\end{document}